\documentstyle[11pt,newpasp,twoside,epsf]{article}
\def\beq{\begin{equation}}
\def\eeq{\end{equation}}

\def\cm{\,{\rm {cm}}}

\def\kms{\,{\rm {km\, s^{-1}}}}

\def\vcir{V_{\rm c}}
\def\v200{V_{200}}

\def\E51{\,{\rm E_{51}}}

\def\S13{S_{-13}}

\def\edcomment#1{\iffalse\marginpar{\raggedright\sl#1\/}\else\relax\fi}
\reversemarginpar

\begin{document}
\title{Physical properties of DLA metallicity and neutral hydrogen column density}
\author{J.L. Hou}
\affil{Shanghai Astronomical Observatory, CAS, Shanghai, China}
\author{C.G. Shu$^{1,2}$, W.P. Chen$^2$, R.X. Chang$^1$ \& C.Q. Fu$^1$}
\affil{$^1$Shanghai Astronomical Observatory, CAS, Shanghai,
China} \affil{$^2$Graduate Institute of Astronomy, National
Central University, Taiwan, China}

\begin{abstract}
We investigate some basic properties of Damped Lyman alpha systems
based on the Semi-Analytical model of disk galaxy formation
theory. We derive the DLA metallicity, column density, number
density, gas content and cosmic star formation rate by assuming
that disks form at the center of dark halos, and the modelled DLAs
are selected by Monte Carlo simulation according to the
distributions of halo properties. We find that DLA hosts are
dominated by small galaxies and biased to extended galaxies. In
terms of model results, DLAs could naturally arise in a
$\Lambda$CDM universe from radiatively cooled gas in dark matter
halos. However, model predicts a reverse correlation between
metallicity and the column density when compared with
observations, regardless of the proposed observational bias. We
argue that this could be resulted from the model limitations, or
the inadequacy of Schmidt-type star formation mode at high
redshift, or/and the diversities of DLA populations.
\end{abstract}

\section{Introduction}

Damped Lyman-$\alpha$ systems (DLAs) are absorbers seen in quasar
optical spectra with HI column density $N_{\rm HI} \ge 10^{20.3}
\cm^{-2}$. DLAs are believed to arise in luminous galaxies or
their progenitors at high redshift. But substantial debate
continues over exactly what populations of galaxies are
responsible for them. Current results of searches appear to
suggest that galaxies giving rise to high HI column density
absorbers span a wide range of morphology types, from dwarf,
irregular, and low surface brightness (LSB) to normal spirals (eg.
Rao et al. 2003). It is also suggested that DLAs could be
associated with Lyman Break Galaxies (LBGs)(Shu 2000; Schaye
2001).

In theory, one way for DLAs research is to assume that DLAs are
galaxies with different morphological types, then the model
predictions can be compared with observed properties (eg. Hou,
Boissier, \& Prantzos 2001; Calura, Matteucci, \& Vladilo 2003;
Boissier, P\'eroux, \& Pettini 2003; Lanfranchi \& Friaca 2003).
Another approach is to start from the framework of cosmic
structure formation and evolution. Hence the observed DLA
properties are strong tests for various cosmological models and
also for Semi-Analytical Model (SAM) of galaxy formation and
evolution (eg. Nagamine, Springel, \& Hernquist 2003; Cora et al.
2003).

Here, we will adopt a SAM method to examine the observed
metallicity and HI column density properties of DLAs. As an
illustration, DLA properties are assumed at redshift $z = 3$ and
the standard $\Lambda$CDM cosmogony is adopted.

\section{The model}

\subsection{Galaxy formation}

The galaxy formation model comes from Mo, Mao, \& White (1998,
hereafter MMW). Detailed descriptions are given in Hou et al.
(2003).

In the model of MMW, the halo mass function at any redshift $z$
can be described by the PS formalism. The distribution function of
halo spin parameter can be described by a log-normal function.
Disks are assumed to have exponential surface profiles $ \Sigma
(R) = \Sigma_0 \exp (-R/R_d)$, where $\Sigma_0$ and $R_d$ are the
central surface density and the scale length. The disk global
properties can be uniquely determined by parameters $m_d$,
$\lambda$, $\vcir$, and the adopted cosmogony, where $m_d$ is the
mass ratio of disk to halo, $\lambda$ is the halo dimensionless
spin parameter, $\vcir$ is the halo circular velocity.

After knowing the distributions of $\vcir$ and $\lambda$ for
halos, we are able to generate a sample of disk galaxies by a
Monte-Carlo simulation in the $\vcir$-$\lambda$ plane at redshift
$z\sim 3$, which is the parent sample for our follow-up DLA
simulation.

\subsection{Star formation and chemical evolution}

The adopted star formation prescription comes from disk galaxy
modelling of Boissier \& Prantzos (2000), in which $ SFR \propto
\Sigma_g^{1.5} V_{rot}R^{-1} $, where $\Sigma_g(R)$ is the gas
surface density and $V_{rot}(R)$ is the rotation speed at disk
radius $R$, which is calculated by considering both components of
halo and disk.

The chemical evolution in disks can be expressed by the simple
closed-box model with metallicity $Z$ to be $Z-Z_i = -p\,{\rm
ln}\mu $, where $Z_i$ is the initial metallicity of gas and is
assumed to be $0.01Z_\odot$, $p$ is the effective yield, and $\mu$
is the gas fraction. We assume that at the initial time ($t = 0,
z=3$), the gas surface density $\Sigma_{g0}(R) = \Sigma_0 \exp
(-R/R_d)$. Star formation proceeds within disks in a typical time
scale $\sim 1 \rm Gyr$ (Lanfranchi \& Friaca 2003). The effective
yield $p$ is assumed to be constant and is obtained by comparing
the metallicity distributions between model predictions and
observations for DLAs.

\subsection{Modelled DLA population}

We investigate star formation and chemical evolution for
individual disk according to the prescriptions mentioned above.
The modelled DLAs are selected over the sampled galaxies by random
sightlines penetrating disks according to the observed selection
criterion. Random inclinations for disks in the sky are considered
and a hydrogen fraction $x = 0.7$ is assumed. We adopt $\vcir$
from 50 to $360\kms$, which corresponds roughly to the observed
range for spirals and irregular galaxies at the present day.

\section{Model prediction vs observations}

\subsection{Observations}

DLAs metallicities are taken from recent compilation presented by
Kulkarni \& Fall (2002), P\'eroux et al. (2002a) and Prochaska et
al. (2003). The observed HI column densities come from the survey
of Storrie-Lombardi \& Wolfe (2000) (hereafter SW00). The observed
number densities and gas content of DLAs come from P\'eroux et al.
(2002). Low redshift gas content comes from Rao \& Turnshek
(2000). The observed contribution to the cosmic SFR by DLAs comes
from the most recent work done by Wolfe, Gawiser, \&
Prochaska(2003) (hereafter WGP03).

\subsection{$\vcir$ and $\lambda$}

\begin{figure} [h]
\plottwo{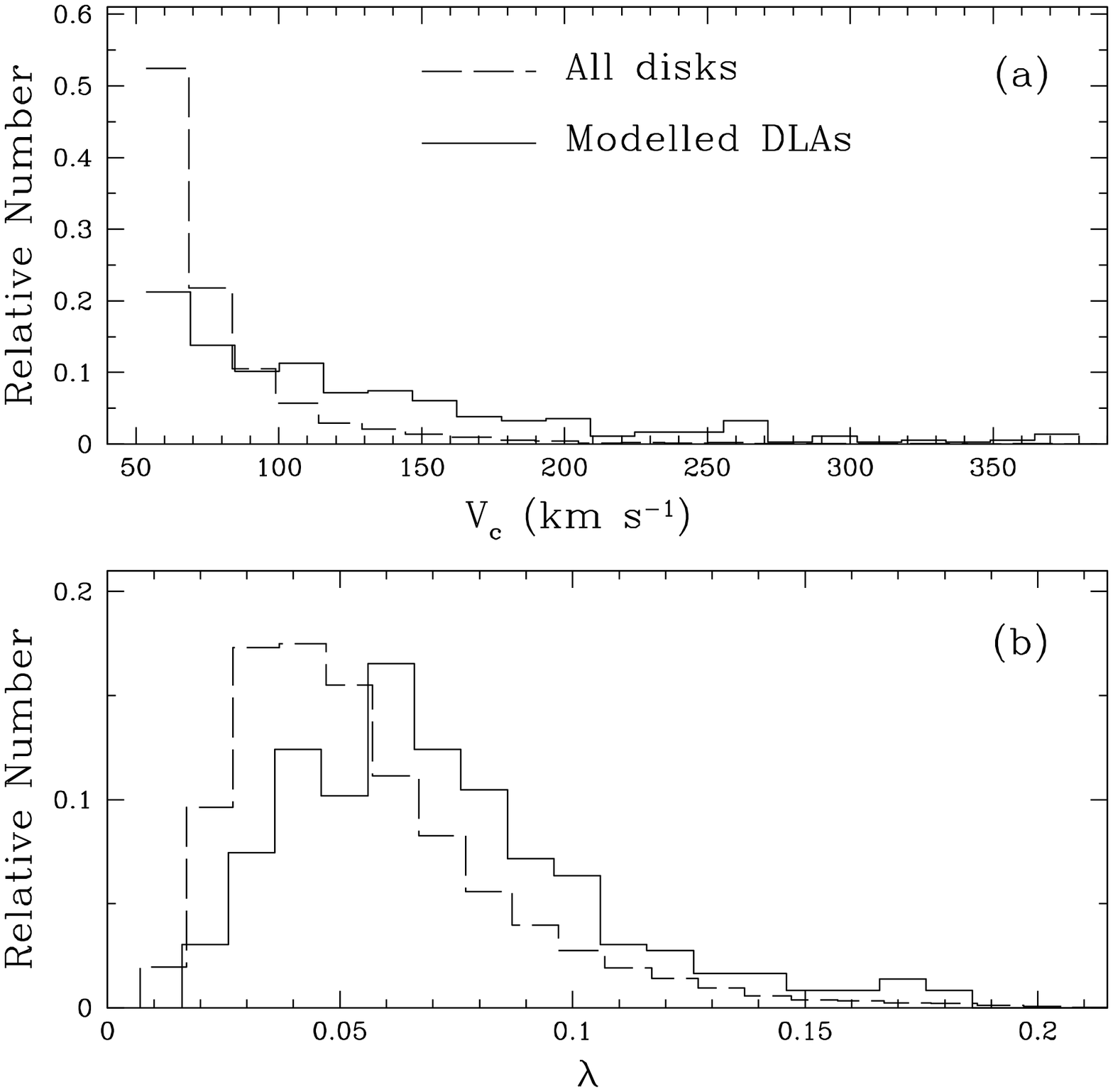}{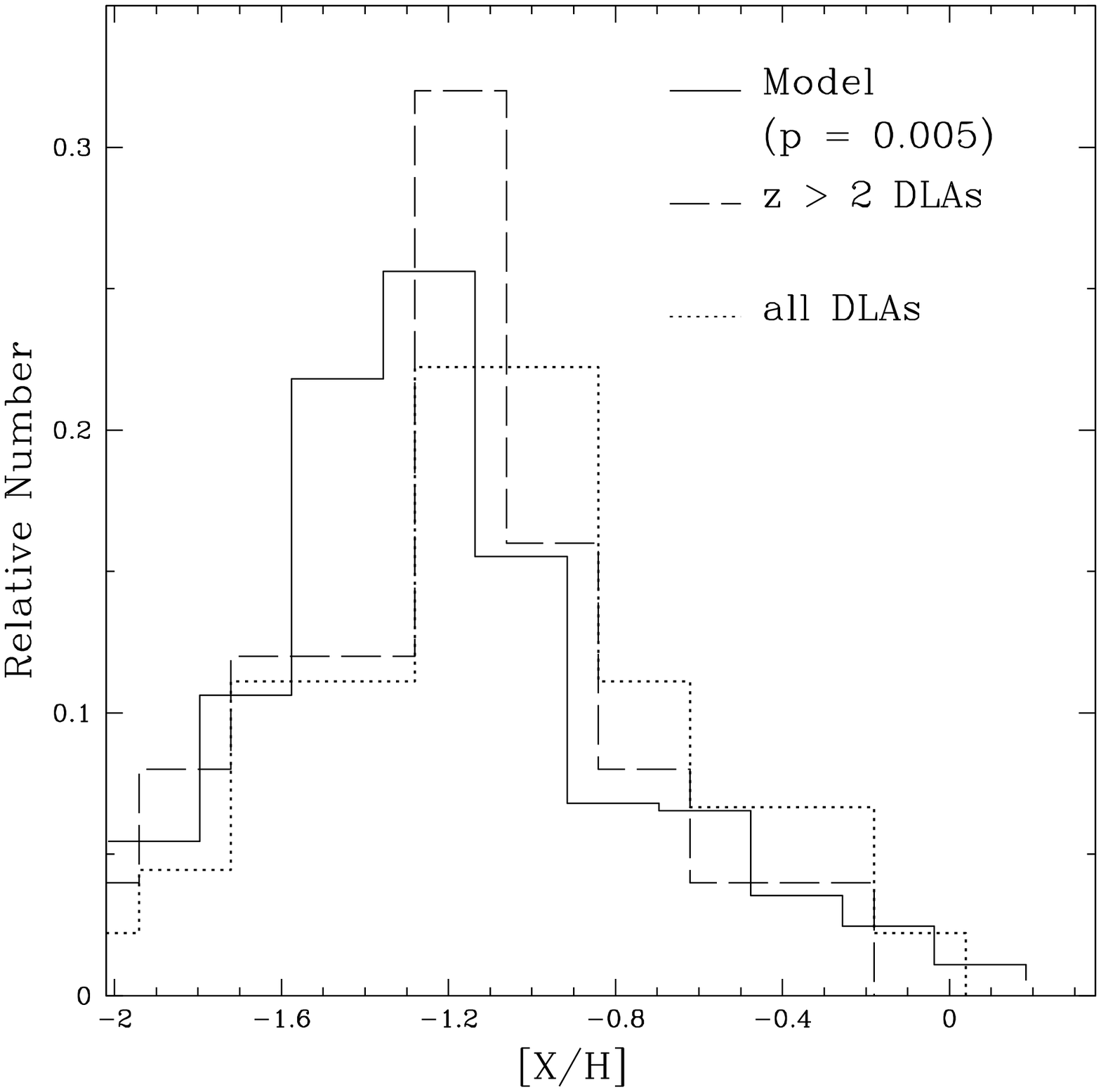} \caption{Left: the
distributions of $\vcir$ and $\lambda$.
Right: the metallicity distributions of DLAs.
} \label{Fig1}
\end{figure}

In the left of Fig. 1, we present the predicted distributions of
$(\vcir, \lambda)$ for the modelled DLA population. We also plot
the corresponding distributions of all galaxies predicted by PS
theory(dash line). It can be found as expected that modelled DLAs
are dominated by small galaxies with $\vcir < 100\kms$. The
$\lambda$ distribution of selected DLAs peaks around 0.065 with
the median value of 0.08, larger than those for all galaxies. DLA
hosts are extended disk galaxies.

\subsection{Metallicity distribution and effective yield}


The predicted and observed metallicity distributions of DLAs are
also plotted in Fig. 1(right). It should be noted that the
predicted distribution is obtained by adjusting the effective
yield $p$. We get $p = 0.25Z_\odot$ for the best-fit between model
and observed [Zn/H] metallicity of $z>2$ DLAs.

The obtained effective yield $p = 0.25 Z_{\odot}$ is low, compared
with that for the solar neighborhood (about 0.7$Z_\odot$). This is
an indication of galactic winds and mass outflows during the
formation of disk from dark halo (Shu, Mo, \& Mao 2003).

\subsection{Column density, number density and gas content}

\begin{figure} [h]
\plottwo{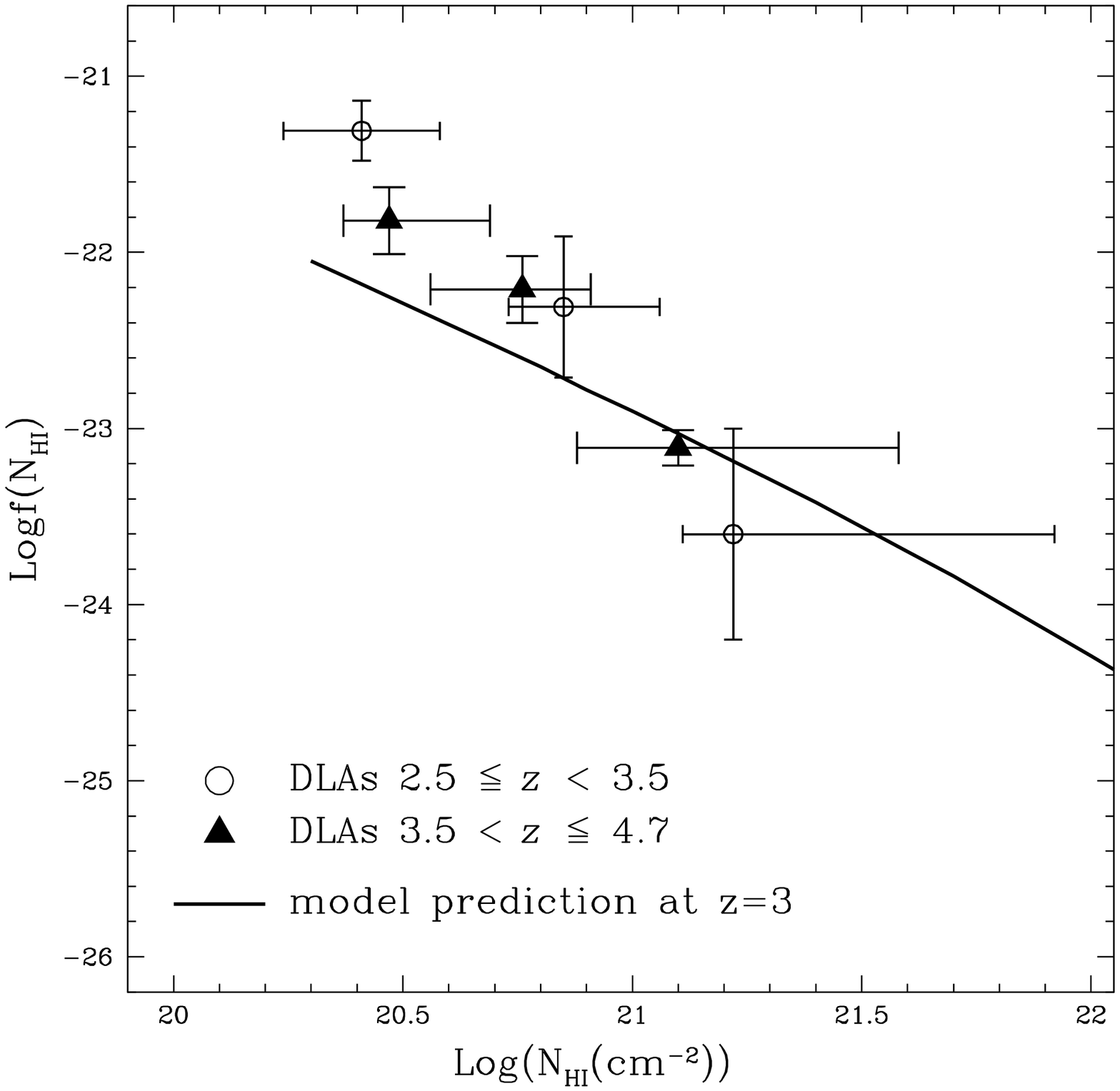}{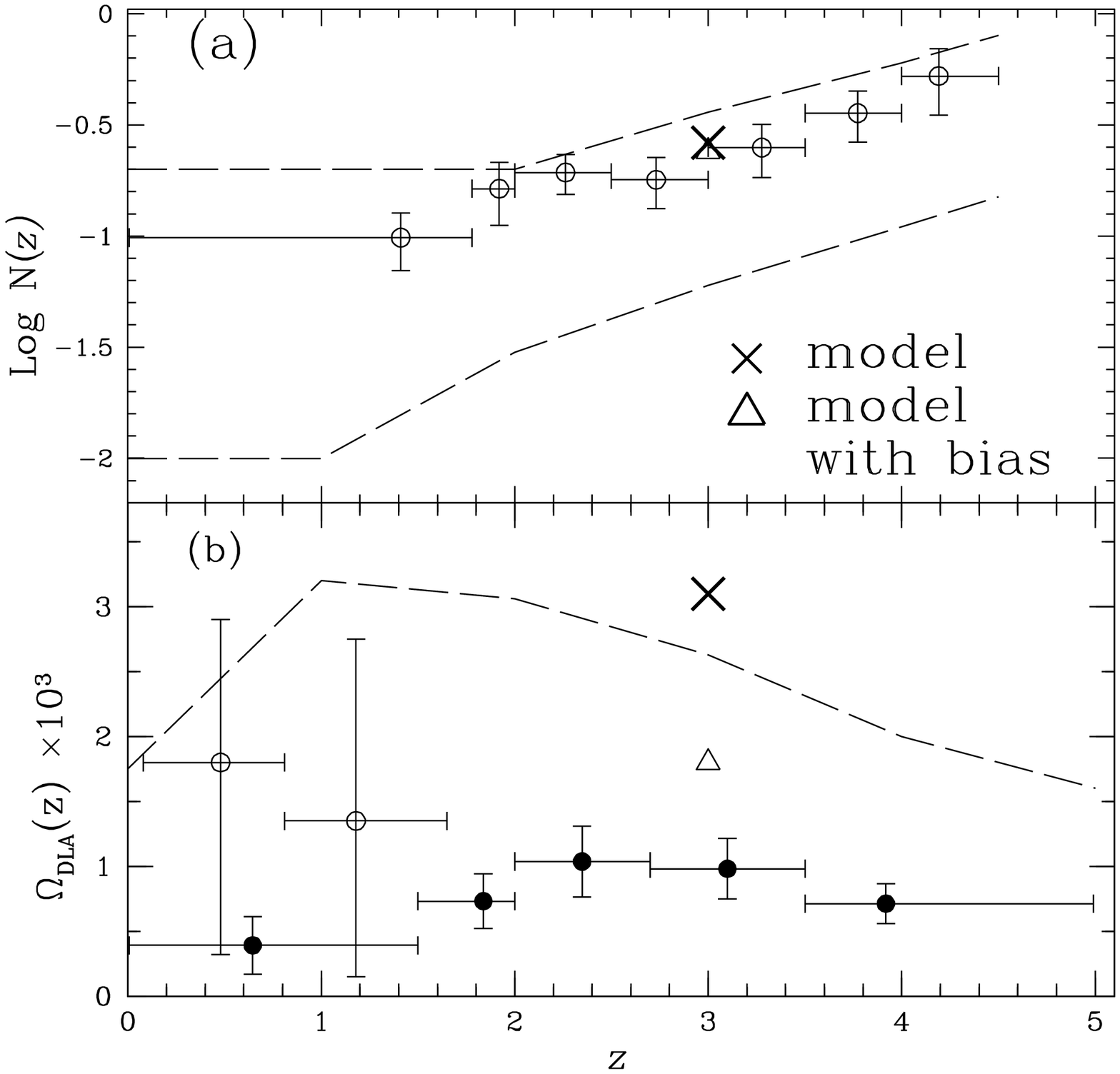} \caption{Left: $f(N_{\rm
HI})$ vs $N_{\rm HI}$ for DLAs. Right: (a) $N(z)$ vs $z$. The two
dash lines denote the lower and upper limits from numerical
simulations (Nagamine et al. 2003); (b) $\Omega_{\rm DLA}$ vs $z$.
The dashed line is the numerical results from Cen et al. (2002).}
\label{Fig2}%
\end{figure}

In Fig.2, we plot the model predicted frequency distribution of
column density $f(N_{HI})$ (left, solid line), number density
$N(z)$ and associated gas content $\Omega_{\rm DLA}$ (right,
cross) at $z = 3$. It can be found from the figure that the
predicted $f(N_{HI})$ distribution agrees well with the high
$N_{\rm HI}$ points, but not with the low $N_{\rm HI}$ values.
This could result from either the model limitations or
observational bias, or both. For instance, galactic winds and mass
outflows could be important for DLAs as implied by the relative
low effective yield obtained. So the closed-box model could be too
relaxed.


In lower panel of right figure (b), the triangle is the predicted
DLA gas content with the consideration of the limitation suggested
by Boiss\'e et al. (1998). In general, we can find that our model
can well reproduce the observed DLA number density at $z\sim 3$,
although the predicted neutral gas density associated with DLAs is
a bit higher than that observed because the upper limit of HI
column density $N_u$ is larger in our modelled DLAs.

\subsection{Cosmic SFR density and metallicity vs column density}

Recently WGP03 have presented SFR density contributed by DLAs
based on the CII absorption lines, which is shown in the left
panel of Fig. 3 as two triangles. Our model prediction at $z = 3$
is plotted as a cross. It can be found that the model prediction
is consistent with observations and supports the "consensus" model
described by WGP03.

\begin{figure} [t]
\plottwo{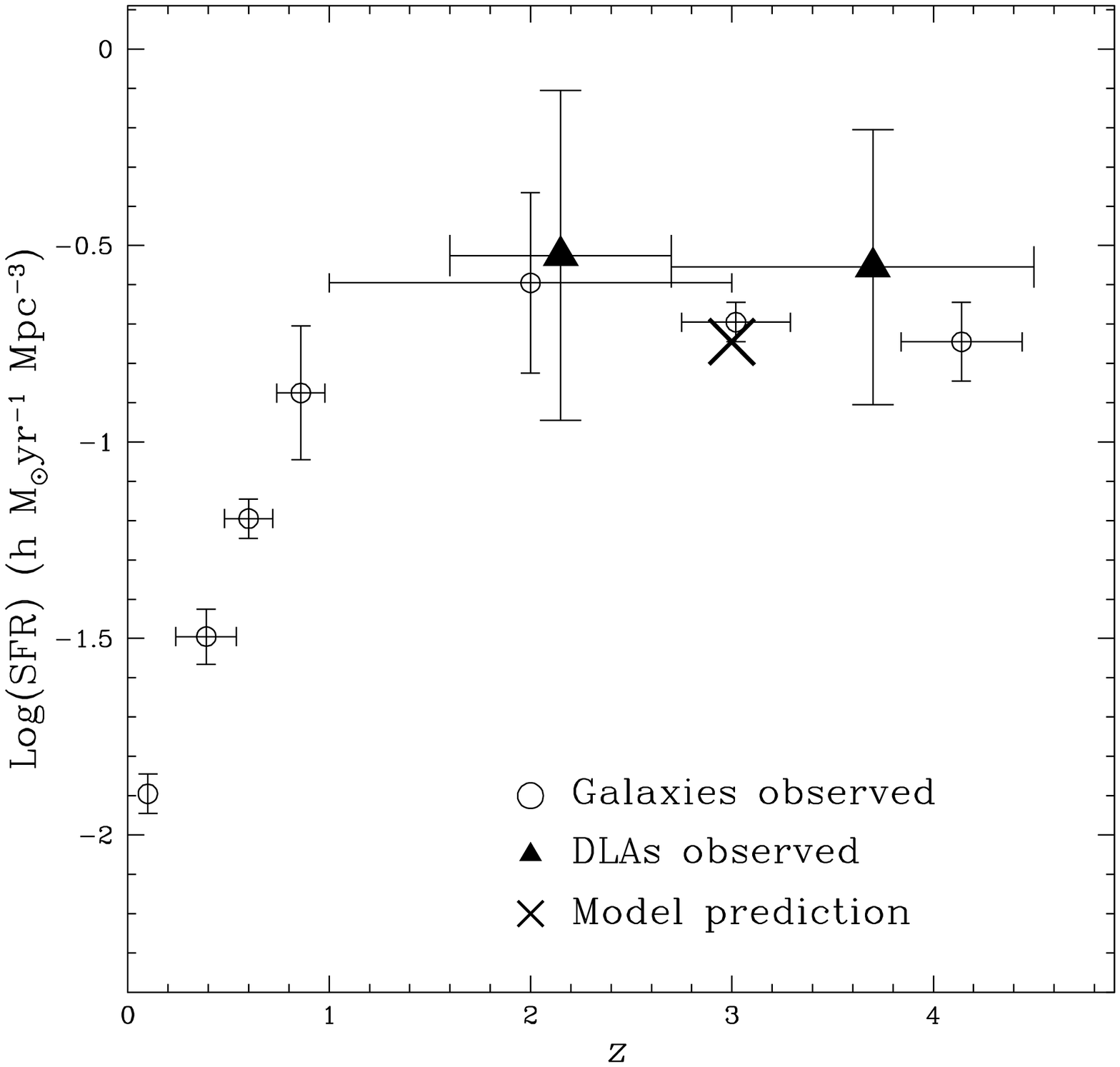}{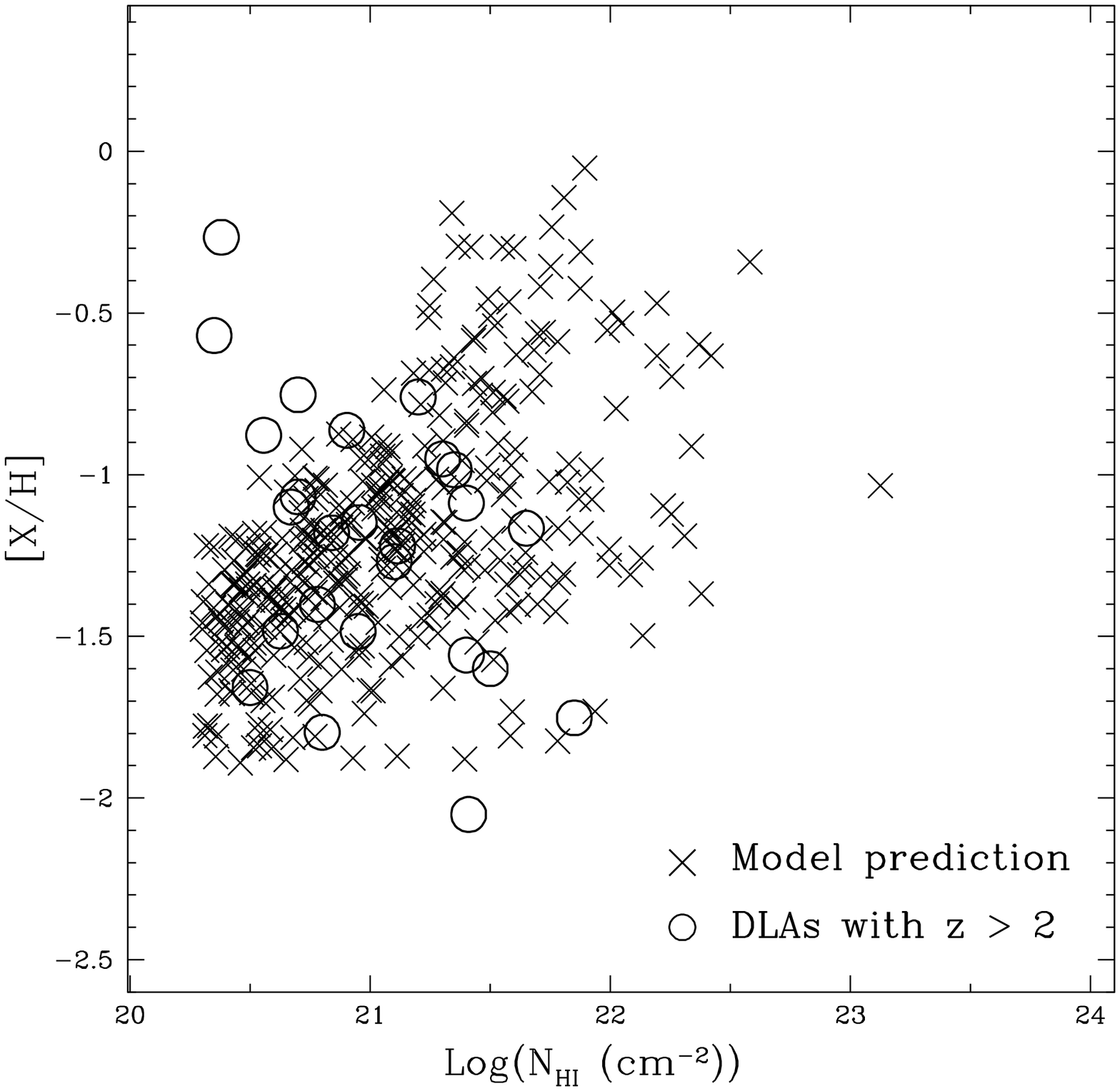} \caption{Right: cosmic SFR
density as a function $z$. Left: The correlation between
metallicity and HI column density for DLAs. }
\label{Fig5}%
\end{figure}


A significant property of DLAs is that there seems to be an
anti-correlation between observed [Zn/H] and HI column density.
However, as given in the right of Fig. 3, model predictions show
an opposite trend compared with observations.

Several mechanisms could lead to the observed anti-correlation.
First, some DLA lines of sight could penetrate galactic central
regions where HI column densities are low and metallicities are
high. Observations of some spiral galaxies have approved this
(Broeils \& van Woerden 1994). However, the absorption cross
sections, hence the probability, are low for this mechanism.

Another mechanism could be the inadequacy of the adopted Schmidt
type star formation prescription. Even for nearby galaxies, the
physical basis of star formation is still poorly known (Kennicutt
1998; Rownd \& Young 1999). Moreover, recent observations of star
formation regions in nearby galaxies done by Wong \& Blitz (2002)
showed a complex relationship between SFR and $\Sigma_{\rm HI}$,
and appears inconsistent with a Schmidt type law.

It is also possible that some observed points resulted from
different populations of DLA galaxies. Observations suggested that
DLAs could be hosted by either disks, spheroids or moving clouds
within galactic halos (Maller et al. 2003). Star formation and
chemical enrichment are quite different for different kinds of
hosts, which could lead to the observed trend in Fig. 3.

\section{Summary}

We have generated a population of disk galaxies by SAM in the
framework standard $\Lambda$CDM hierarchical picture of structure
formation. Modelled DLAs are selected according to their
observational criterion with the random inclination being
considered.

With the effective yield $p = 0.25Z_{\odot}$ obtained for the
corresponding best-fit result, our model can well reproduce the
observed metallicity distribution of DLAs.

In terms of predicted results of the HI frequency distribution,
the number density, gas content and cosmic SFR density at redshift
3, our model suggests that DLAs could naturally arise in a
$\Lambda$CDM universe from radiatively cooled gas in dark matter
halos.

Model predicts a positive correlation between metallicity and HI
column density for DLAs, inconsistent with the observed trend. We
suggest that the observed anti-correlation could most probably be
physical.

\acknowledgments{We are grateful to B. M\'enard, S. Boissier, W.P.
Lin, H.J. Mo, D.H. Zhao for useful discussions on this subject.
This work is supported by NSFC19873014, 10073016, 10173017,
10133020, NKBRSF 19990754 and NSC91-2112-M008-036. CS acknowledges
the financial support of NSC for a visit to NCU and the kind
hospitality during the stay in NCU.

\def\cjaa{CJAA}
\def\aapss{AApSS}

\end{document}